\documentclass[aps,prl,twocolumn,showpacs,superscriptaddress,groupedaddress]{revtex4}  

\usepackage [dvips]{graphicx}
\usepackage{graphicx}  
\usepackage{dcolumn}   
\usepackage{bm}        
\usepackage{amssymb}   
\usepackage{amsbsy}    
\usepackage{color}

\usepackage{amsmath}
\usepackage{epsfig}
\usepackage{verbatim}

\definecolor{orange}{rgb}{1.0,0.75,0}

\addtocounter{secnumdepth}{1}
\font\capfont=cmbx12 at 50 pt 
\newbox\capbox \newcount\capl \def\a{A}
\def\docappar{\medbreak\noindent\setbox\capbox\hbox{%
\capfont\a\hskip0.15em}\hangindent=\wd\capbox%
\capl=\ht\capbox\divide\capl by\baselineskip\advance\capl by1%
\hangafter=-\capl%
\hbox{\vbox to8pt{\hbox to0pt{\hss\box\capbox}\vss}}}
\def\cappar{\afterassignment\docappar\noexpand\let\a }

\begin{document}

\newcommand{\p}{P}
\newcommand{\pc}{{\cal P}}
\newcommand{\deltapc}{{\cal Q}}
\newcommand{\pdw}{P_{MF}}
\newcommand{\pcdw}{{\cal P}_{MF}}
\newcommand{\deltadw}{\delta_{MF}}
\newcommand{\deltame}{\delta}
\newcommand{\Dp}{D_{+}}
\newcommand{\Dm}{D_{-}}
\newcommand{\Dpm}{D_{\pm}}
\newcommand{\Dmp}{D_{\mp}}
\newcommand{\Ddw}{D_{MF}}
\newcommand{\Dmeo}{D_{1}}
\newcommand{\Dmet}{D_{2}}
\newcommand{\Dme}{D}
\newcommand{\rhop}{\rho_{+}}
\newcommand{\rhom}{\rho_{-}}
\newcommand{\rhopm}{\rho_{\pm}}
\newcommand{\rhomp}{\rho_{\mp}}
\newcommand{\rhodw}{\rho_{MF}}
\newcommand{\jp}{j_{+}}
\newcommand{\jm}{j_{-}}
\newcommand{\jpm}{j_{\pm}}
\newcommand{\jmp}{j_{\mp}}
\newcommand{\xidw}{\xi_{MF}}
\newcommand{\xime}{\xi}
\newcommand{\dt}{\partial_t}
\newcommand{\deltat}{\Delta_t}
\newcommand{\dtau}{\partial_\tau}
\newcommand{\dx}{\partial_x}

\newcommand{\deltapcz}{\deltapc^{(0)}}
\newcommand{\deltapco}{\deltapc^{(1)}}
\newcommand{\deltapct}{\deltapc^{(2)}}
\newcommand{\stat}{\mathrm{stat}}
\newcommand{\DWT}{DWT }
\newcommand{\SDW}{SDW }
\newcommand{\SDWT}{SDWT }
\newcommand{\FDP}{FDP}
\newcommand{\FDPs}{{\FDP}s }

\newcommand{\la}{\langle}
\newcommand{\ra}{\rangle}
\newcommand{\beq}{\begin{equation}}
\newcommand{\eeq}{\end{equation}}
\newcommand{\bea}{\begin{eqnarray}}
\newcommand{\eea}{\end{eqnarray}}
\def\lsim{\:\raisebox{-0.5ex}{$\stackrel{\textstyle<}{\sim}$}\:}
\def\gsim{\:\raisebox{-0.5ex}{$\stackrel{\textstyle>}{\sim}$}\:}



\title{Exact domain wall theory for deterministic TASEP with parallel update}

\author{{ J. Cividini, H.J.~Hilhorst, and C.~Appert-Rolland}\\[2mm]
{\small Laboratoire de Physique Th\'eorique, b\^atiment 210}\\
{\small Universit\'e Paris-Sud and CNRS (UMR 8627),
91405 Orsay Cedex, France}\\}

\begin{abstract}
Domain wall theory (DWT) has proved to be a powerful
tool for the analysis of one-dimensional transport processes. 
A simple version of it was found very accurate for the Totally Asymmetric
Simple Exclusion Process (TASEP) with random sequential update. However, a general implementation of DWT is still missing in the case of updates with less fluctuations, which are often more relevant for applications.
Here we develop an exact DWT
for TASEP with parallel update and
deterministic ($p=1$) bulk motion. 
Remarkably, the dynamics of this system can be described by the motion of a domain wall not only on the coarse-grained level but also exactly on the microscopic scale for arbitrary system size.
All properties of this TASEP, time-dependent and stationary, 
are shown to follow from the solution of a \textit{bivariate} master equation
whose variables are not only the position but also the velocity of the
domain wall.
 In the continuum limit 
this exactly soluble model then allows us to perform 
a first principle derivation of a Fokker-Planck equation for the
position of the wall. 
The diffusion constant appearing in this equation differs from the one
obtained with the
traditional ``simple'' DWT.
\end{abstract}


\maketitle

Nature abounds with problems controlled by unidirectional
one-dimensional transport. The transport may be in channels 
(as in porous materials or across cell membranes) or along
rails 
(\textit{e.g.} the cytoskeleton of biological cells).
The analogy to road and pedestrian traffic 
has reinforced interest in such systems and spurred research aimed
at uncovering common characteristics.

\begin{figure}
\begin{center}
\scalebox{.4}
{\includegraphics{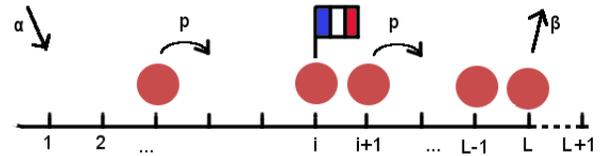}} 
\end{center}
\caption{\small TASEP of sites $1,2,\ldots,L$. 
Red disks represent particles and arrows represent transition
probabilities for the particles; this work studies the case $p=1$.	
A particle carrying a `flag' separates a low from a high
density phase (see text).}  
\label{fig_scheme}
\end{figure}

Several simple models have been proposed for the
description of such phenomena. Among these 
the Totally Asymmetric Simple Exclusion Process (TASEP) has become a
major paradigm of out of equilibrium systems. In this model
each site $1,2,\ldots,L$ of a finite one-dimensional lattice (see
Fig.\,\ref{fig_scheme}) is either empty or singly occupied.
Particles are injected onto site $1$, may
hop to the right if their target site is empty,
and are removed from site $L$. 
The order in which these steps are performed (the `updating scheme')
completes the definition of a specific TASEP.

The random sequential update~\cite{derrida1998a,chou_m_z2011} often
used for the TASEP leads to large fluctuations in individual velocities.
In real road or pedestrian traffic, particle motion tends to be synchronous. 
Therefore TASEP based traffic modeling~\cite{chowdhury_s_s2000}  
rather uses parallel update~:
the configuration at time $t+1$
is obtained from the one at time $t$ by moving 
with probability $p$ each particle
 with an empty target site one step to its right; 
filling the leftmost site 1 with probability $\alpha$ if empty; and 
removing the particle on site $L$, if any, with probability $\beta$.

Exact results on TASEPs, in particular concerning their stationary
states~\cite{schutz_d1993,derrida1993c,evans_r_s1999}, are known for various
updating schemes but require considerable mathematical
sophistication.  
In order to obtain results that are beyond the scope of exact solutions, Kolomeisky {\it et al.}~\cite{kolomeisky1998} have applied the more general phenomenological approach of \textit{domain wall theory} (DWT) to TASEP. Their implementation, to which we shall refer as \textit{simple} DWT (SDWT), successfully predicts dynamical
quantities of the TASEP when its update is random sequential~\cite{dudzinski_s2000,nagy_a_s2002,pierobon2005} and can be easily
adapted to variants with modified
kinetics~\cite{juhasz_s2004,reichenbach_f_f2006,reichenbach_f_f2008,parmeggiani_f_f2004} and
geometries~\cite{schiffmann_a_s2010,arita_s2011a,pronina_k2005,yuan2008b},
or be used as a basis for more general
discussions~\cite{popkov_s1999}.
For other updates, however, observed discrepancies with exact~\cite{pigorsch_s2000,arita_s2011a} or
numerical~\cite{appert-rolland_c_h2011b, cook_z2009} results call for an adapted DWT such as proposed in Ref.~\cite{belitsky_s2011b,pigorsch_s2000} for sublattice parallel update.

The purpose of this Letter is to
build a complete and exact (and not only phenomenological) DWT
for the TASEP with parallel update and $p=1$. 
The detailed study of this model was initiated by
Tilstra and Ernst \cite{tilstra_e1998} before the advent of DWT.
Here we show that there exists a pair of domain wall variables 
(to be called the `flag position' and the `flag velocity') 
that satisfies an exact master equation.
We then derive a DWT from first principles, which
appears to contain a diffusion constant $D$ different from the one in SDWT.
We shall first summarize the SDWT approach.
\vspace{2mm}

DWT approximates the system by two spatially uniform domains separated by a domain wall. On the right of the wall,
the queue
of particles that have been blocked at the exit
constitutes a jammed phase, while on the left a free flow domain
is sustained by the entrance boundary.
This description applies for $\alpha$ and $\beta$ below a critical value
(above which a maximum current phase may appear, which we do not consider here).
The domain wall, assumed to be of
negligible width, is located at a position $i$ that fluctuates with time. 
Let $\rhopm$ be the average particle densities of the right and left domain,
respectively, and $\jpm$ the corresponding currents, all supposed known. The first
postulate of SDWT, as applied by Ref.~\cite{kolomeisky1998} and tested
by Ref.~\cite{santen_a2002},  is that
the probability $\p_i(t)$ to find the domain wall on
$i$ at time $t$ satisfies the master equation 
\beq
\frac{d\p_i(t)}{dt} = \Dp \p_{i-1}(t) + \Dm \p_{i+1}(t) - (\Dp + \Dm) \p_i(t)
\label{eq_medw}
\eeq
with reflecting boundary conditions at $i=1$ and $i=L$. 
If this equation is true at all, it should be possible to express the
{\it a priori\,} unknown coefficients $\Dpm$ in terms of the basic
model parameters.
Mass conservation imposes 
$ \Dp - \Dm = \frac{\jp-\jm}{\rhop-\rhom}$. However, 
as rightly pointed out by 
Kolomeisky {\it et al.} \cite{kolomeisky1998}, an extra
hypothesis concerning the current fluctuations is needed
to determine $D\equiv(\Dp+\Dm)/2$, which in the continuum
limit becomes the diffusion constant of a Fokker-Planck equation.  
Observing that for TASEP $\jpm=0$ implies $\Dpm=0$ the second postulate of SDWT is that
$\Dpm = \jpm/(\rhop-\rhom)$, so that $D = \frac{1}{2}\frac{\jp+\jm}{\rhop-\rhom}$.
For the random sequential update, the SDWT expressions for $\Dpm$ are supported by the fact that, for large systems, they reproduce correctly the exactly known~\cite{schutz_d1993,derrida1993c} stationary density profile.
Besides, in the case of random sequential update many TASEP properties, both dynamical and
stationary, are reproduced accurately by
SDWT~\cite{kolomeisky1998,santen_a2002}.
However, as mentioned, SDWT is not appropriate for updates
with low fluctuations, and we shall now derive a complete DWT
in the case of the
deterministic parallel update.
In contrast to SDWT, this derivation is exact and
thus no postulates are needed.

The $p=1$ TASEP with parallel update has completely deterministic
bulk dynamics; 
only the entrance and exit processes are stochastic. 
Two domains may coexist in this model, a free flow domain
on the left and a jammed domain on the right.
In the free flow domain particles enter randomly at the left and advance
at unit velocity; two successive particles
are separated by at least one hole. The site occupation probabilities
$\rho_i(t)$ then satisfy simple recursion relations in space: if $i$
is occupied, $i-1$ is empty; and if $i$ is empty, $i-1$ is
occupied with probability $\alpha$. The jammed domain has a symmetrical structure obtained by exchanging particles and holes, $\alpha$ and $\beta$, and right and left.
The domains have~\cite{rajewsky1998} 
\beq
\jm = \rhom = \alpha/(1+\alpha), \quad \jp = \beta \rhop = \beta/(1+\beta).
\label{eq_jrhop}
\eeq
For $\alpha < \beta$ the system is in free flow phase, \textit{i.e.} the free flow domain invades the bulk, while for $\alpha > \beta$ the system is in the jammed phase. The critical line is $\alpha = \beta$.

The system contains 
at any instant of time two classes of particles, {\it viz.} on the
left those of the free flow domain, that have never been blocked, and on the right those of the jammed domain,
 that have been blocked at least once during their travel through the system. 
We will say that the leftmost particle ever to have been blocked 
carries a {\it flag}
\footnote{The flag definition, which is with
  respect to the particles, breaks 
  the particle-hole symmetry. A similar flag could be defined with
  respect to the holes and a symmetric description would consider the
  joint distribution of the variables of the two flags.}
similar to the shock marker introduced in Ref.~\cite{belitsky_s2011a}.
If no particle in the system has ever been blocked, the flag occupies
by convention a virtual site $L+1$. A typical configuration is
depicted in Fig.\,\ref{fig_scheme}.
A particle can get blocked only if its predecessor has been blocked,
so that by induction all particles to the right of the flag have also
undergone blocking. 
\vspace{2mm}

We are now interested in the time evolution of the probability distribution 
$\p_i(t)$ of the flag position. At each time step the flag may execute hops 
$i\to i,i\pm 1$ according to the rules below.
When the flag carrying particle blocks the forward move of a particle
to its left, the flag is transferred to this latter particle, that is,
hops one lattice distance to the left. 
In the other cases the flag remains attached to
its carrier particle which may either hop forward
or stay on the same site.
It can be seen
that the random motion of the flag has a memory
of one time step. Indeed, let us define $\p^a_i (t)$, for $a=0,\pm 1$, 
as the probability that at time $t$ the flag is on site $i$ 
{\it and\,} has arrived there by a move of 
$a$ lattice units in the preceding time step.
Hence $a$ may be interpreted as
 the flag velocity between $t-1$ and $t$. 

Conditional on the flag having arrived at site $i$ with velocity $a$ 
we know the following.
Site $i$ is occupied by a particle for sure and
$i+1$ is occupied with probability $1-\beta$.
For $a=1$ site $i-1$ is empty and for $a=0,-1$ it is occupied with
probability $\alpha$. 
This knowledge determines the probabilities for what will happen at
the next time step. 
If we set $\p^{1}_1(t)\equiv 0$ and $\p^{-1}_{L+1}(t)\equiv 0$,
we may take the cases $a=0,-1$ together by introducing $\p^{-10}_i(t)
\equiv \p^{-1}_i(t) + \p^{0}_i(t)$ for $1\leq i\leq L+1$. 
The equations for $\p^{-10}$ and $\p^{1}$ then read
\bea
\label{eq_me2}
\p^{-10}_i(t+1) & = & \alpha \p^{-10}_{i+1}(t) + (1-\alpha)(1-\beta)
\p^{-10}_{i} (t) \nonumber \\ 
&&+\, (1-\beta) \p^{1}_{i} (t),  \nonumber\\
\p^{1}_i(t+1) & = & (1-\alpha)\beta \p^{-10}_{i-1}(t) + \beta \p^{1}_{i-1}(t), 
\eea
for $2\leq i\leq L-1$ and $3\leq i\leq L$, respectively. 
Near the left boundary we have the two special equations
\bea
\label{eq_bc2_left}
\p^{-10}_1(t+1) & = & \alpha \p^{-10}_2(t) + (1-\beta) \p^{-10}_1(t),
\nonumber \\ 
\p^{1}_2(t+1) & = & \beta \p^{-10}_1(t).
\eea

The right boundary requires more attention. 
If the flag has just arrived on site $L+1$ ({\it i.e.} has $a=1$),
then site $L$ is occupied with probability $r_0=0$. But if the
flag stays on $L+1$ (has $a=0$), this probability evolves with each
time step. Let $r_u$ be the occupation probability of site $L$
after the flag has stayed on $L+1$  
for $u$ time steps.
The $r_u$ can be calculated from an elementary
recursion in $u$. 
In order to accommodate this memory effect at the right boundary into
a Markovian description, 
we write $ \p^{-10}_{L+1}(t) = \sum_{u=1}^\infty \p^{-10}_{L+1,u}(t)$
in which $\p^{-10}_{L+1,u}(t)$ takes into account the time $u$ the
flag has spent on site $L+1$ since its latest arrival there. 
The special evolution equations near the right boundary then read
\bea
\label{eq_bc2_right}
\p^{-10}_{L} (t+1) &=& \sum_{u=1}^\infty (1-\beta) r_u
\p^{-10}_{L+1,u} (t)  \nonumber \\ 
&& + (1-\alpha)(1-\beta) \p^{-10}_{L} (t) + (1-\beta) \p^{1}_{L} (t),
\nonumber \\ 
\p^{-10}_{L+1,1}(t+1) &=& \p^1_{L+1}(t),  \\
\p^{-10}_{L+1,u}(t+1) &=& (1-(1-\beta) r_{u-1}) \p^{-10}_{L+1,u-1}(t),
\quad u \geq 2. \nonumber 
\eea
The closed system of equations
(\ref{eq_me2})-(\ref{eq_bc2_right}), valid for all $L \geq 2$ 
  constitutes the master
  equation of our `flag theory'.

In the stationary state we have
\beq
\p^{-10,\stat}_i = \beta^{-1}\p^{1,\stat}_{i+1} = Z^{-1}(\beta/\alpha)^{i-1}
\label{eq_mestat}
\eeq
for $1\leq i\leq L$, together with $\p^{1,\stat}_1 = 0 $ and
$\p^{-10,\stat}_{L+1} = \frac{1}{Z} \frac{1+\alpha\beta}{1-\beta}
\big( \frac{\beta}{\alpha} \big)^L$, 
and where $Z$ is the normalization constant.
It shows that the probability $\p_i^{\stat}$ is concentrated near the
left (the right) boundary when $\alpha>\beta$ (when $\alpha<\beta$),  
within a penetration depth $\xi=|\log(\beta/\alpha)|^{-1}$,
in agreement with the findings of Ref.\,\cite{tilstra_e1998}.
\vspace{1mm}

\begin{figure}
\begin{center}
\scalebox{.32}
{\includegraphics{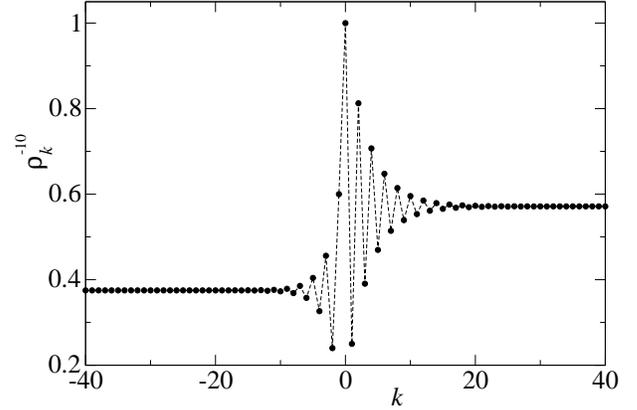}} 
\end{center}
\caption{\small Flag dependent density profile $\rho^{-10}_{k}$ 
 for $\alpha = 0.60$ and
  $\beta = 0.75$.   
For $k\to\mp\infty$ it tends to the bulk values $\rhomp$ as
$\sim(-\alpha)^{|k|}$ on the left and as $\sim(-\beta)^k$ on the
right. 
Dotted lines are guides to the eye.
}
\label{fig_profile}
\end{figure}

Now we wish to calculate the density profile.
Let $\rho^a_{j-i}$ (with $1\leq i,j\leq L$) be the 
expected particle density on site $j$ conditionally on the flag
being at $i$ with velocity $a$.  
We anticipate that this `flag dependent profile' (FDP) depends only on the
difference $j-i \equiv k$. 
The $\rho^a_{k}$ follow from
the appropriate recursion relation in space 
(the ones for the free flow and the jammed phase 
for $k<0$ and $k>0$, respectively);
the knowledge of the flag velocity $a$ at site $i$ provides the starting values.
Fig.\,\ref{fig_profile} shows one of these FDPs.
For $i=L+1$ the special $u$-dependent FDPs may be calculated similarly.

\begin{figure}
\begin{center}
\scalebox{.32}
{\includegraphics{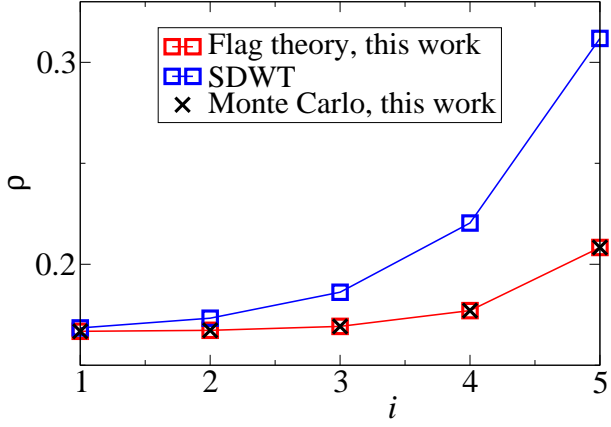}} 
\end{center}
\caption{\small Stationary density profile for $L=5$, $\alpha=0.2$ and
  $\beta=0.8$. 
Open blue squares: SDWT. 
Black crosses and open red squares: Monte Carlo simulation
and the flag theory of this work, respectively.
Lines are guides to the eye.} 
\label{fig_comprho}
\end{figure}

The time-independent profiles $\rho^a_{j-i}$ are 
attached to the frame of reference of the moving flag.
The time dependent density $\rho_j(t)$ at site $j$ is the
average of $\rho^a_{j-i}$ with respect to the distributions $P_i^a(t)$
of the flag position and velocity, with proper account of
contributions from the special flag position at $L+1$.
Fig.\,\ref{fig_comprho} shows that our Monte Carlo results
for the stationary state, even for a small system, $L=5$, are in excellent
agreement with the flag theory, as expected of an exact theory. 
\vspace{2mm}

We now scale the lattice coordinate as $x=i/L$
and consider the limit $L \rightarrow \infty$, adopting the notation
$\p^a_{i}(t) \equiv L^{-1} \pc^a(x,t)$.
For $\alpha,\beta < 1$ the 
FDPs then become step functions as in SDWT. 
We will show that, in the large $L$ limit, it is possible to extract from (\ref{eq_me2})-(\ref{eq_bc2_right}) an
equation for the position distribution of the flag 
$\pc = \pc^{-10}+\pc^1$ alone. 
We introduce the 
shorthand notation $\deltapc \equiv \frac{(1-\beta)
\pc^1 - \beta (1-\alpha) \pc^{-10}}{1-\alpha\beta}$, $\deltat A(t)
\equiv A(t+1) - 
A(t)$ for any quantity $A$, 
and $\deltame \equiv \frac{\beta-\alpha}{1-\alpha\beta}$.
We also define
\beq
\label{eq_D1}
\Dmeo \equiv \frac{1}{2} \frac{\alpha+\beta -2
\alpha\beta}{1-\alpha\beta}.
\eeq 
When Taylor expanding all quantities in Eqs.\,(\ref{eq_me2}) around $x = i/L$
we find 
\begin{subequations}
\label{eq_fp2bis}
\bea
\deltat \pc & = & - \frac{\deltame}{L} \frac{d \pc}{dx}+ \frac{\Dmeo}{L^2}
\frac{d^2 \pc}{dx^2} - \frac{(1+\beta)\alpha}{L} \frac{d \deltapc}{dx} \nonumber \\ 
&& - \frac{(1-\beta)\alpha}{2 L^2} \frac{d^2 \deltapc}{dx^2} + O(L^{-3}),
\label{eq_fp2bisa}
\eea
\bea
\deltat \deltapc & = & - (1-\alpha\beta) \deltapc 
- \frac{\beta(1-\alpha^2)(1-\beta)}{L
  (1-\alpha\beta)^2} \frac{d \pc}{dx} \nonumber \\  
&& + \frac{\alpha\beta(\beta-\alpha)}{L(1-\alpha\beta)} \frac{d \deltapc}{dx} +
O(L^{-2}).  
\label{eq_fp2bisb}
\eea
\end{subequations}
We may solve Eq.\,(\ref{eq_fp2bisb}) for $\deltapc$ in terms of
$\pc$. The term $-(1-\alpha\beta)\deltapc$ in this equation causes
$\deltapc$ to decay to values $\sim L^{-1}$ on a time scale $\sim L^0$
(which was the reason for defining $\deltapc$ as we did) and hence,
on time scales $\gg L^0$,  
\beq
\deltapc = - \frac{\beta(1-\beta)(1-\alpha^2)}{L(1-\alpha\beta)^3} \frac{d \pc}{dx} + O(L^{-2}). 
\label{eq_deltapcstat}
\eeq
Substituting Eq.\,(\ref{eq_deltapcstat}) in Eq.\,(\ref{eq_fp2bisa}) gives
\beq
\label{eq_fppc}
\deltat \pc = - \frac{\deltame}{L} \frac{d \pc}{dx} + \frac{\Dme}{L^2} \frac{d^2 \pc}{dx^2} + O(L^{-3}), 
\eeq
in which the diffusion constant is given by
$\Dme \equiv \Dmeo + \Dmet$ 
where $\Dmeo$ is given by (\ref{eq_D1}) and $\Dmet$ stems from $\deltapc$,
\beq
\label{eq_D2}
\Dmet \equiv \frac{\alpha \beta (1-\alpha^2) (1-\beta^2)}{(1-\alpha\beta)^3}.
\eeq
The constant $D$ differs from the one found by Belitsky \textit{et al.}~\cite{belitsky_s2011a}, which applies to a particular type of shock. It is also different from that of SDWT.

\begin{figure}
\begin{center}
\scalebox{.32}
{\includegraphics{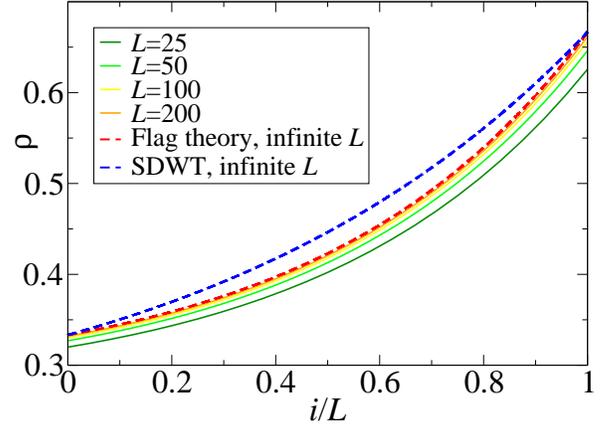}} 
\end{center}
\caption{\small Stationary density profile in the scaling limit for
  $\beta=0.5$ and $c=1$. 
Blue dotted line: SDWT. It is distinctly
different from the
red dashed line, obtained from Eq.\,(\ref{eq_scaling}) and representing
the flag theory of this work.
Continuous lines: Monte-Carlo simulations for different lengths $L$
converging to the flag theory prediction.}  
\label{fig_scaling}
\end{figure}

Whereas Eq.\,(\ref{eq_fppc}) is valid for arbitrary $\alpha,\beta$,
it is of interest near the phase transition line $\alpha=\beta$ to
investigate the scaling limit  
$L\to\infty$ with $\beta-\alpha = c/L$ and the constant $c$ fixed.
Repeating the calculation with
$\tau \equiv t L^{-2}$ we find that in this limit $D=(1-\beta^2)^{-1}$ while SDWT would have given $D = \beta (1-\beta)^{-1}$.
$\pc$ satisfies the Fokker-Planck equation
\beq
\label{eq_scaling}
\dtau \pc = (1-\beta^2)^{-1} \left( -c \dx \pc + \dx^2 \pc \right).
\eeq
The boundary conditions associated with Eq.\,(\ref{eq_scaling}) may be
derived from Eqs.\,(\ref{eq_bc2_left}) and (\ref{eq_bc2_right}) by a
calculation similar to that of Ref.\,\cite{vankampen_o1972}
(Sec.\,5). On the time scale $\tau$ the memory effect at $i=L+1$
collapses and we obtain at both ends of the interval standard zero
current boundary conditions.
Fig.\,\ref{fig_scaling} shows that the
profile obtained by Monte Carlo simulation of a size $L$ system indeed
converges in the scaling limit towards the stationary distribution of
Eq.\,(\ref{eq_scaling}). 
\vspace{2mm}

This work arose from the need to extend the DWT
beyond random sequential update.
Here, in the context of the $p=1$ TASEP
with parallel update, we developed a full DWT
which, in contrast to the SDWT, is exact even at the microscopic scale and for systems of any finite size.
Indeed, the dynamics of this model can be reduced
to a Markov process more complicated than that of SDWT
and involving the position and speed of a `flag'. 
In the continuum limit a
Fokker--Planck equation results, but with a diffusion
constant different from that of SDWT
and in agreement with Monte Carlo simulations.
\vspace{2mm}

The authors thank Kirone Mallick and Chikashi Arita 
for `tea group' discussions that triggered this work.

\end{document}